\documentclass{elsart3p}

\usepackage{amssymb}
\usepackage{graphicx} 

\begin{document}

\def\sn1006{SN\,1006}
\def\rxj1713{RX\,J1713.7$-$3946}
\def\casa{Cassiopeia\,A}
\def\velajr{RX\,J0852.0$-$4622}
\def\rcwa86{RCW\,86}
\def\w28{W\,28}
\def\ic443{IC\,443}

\def\ctba37{CTB37\,A}
\def\ctbb37{CTB37\,B}
\def\ga09p01{G\,0.9+0.1}
\def\gb128m02{G\,12.8$-$0.0}
\def\gc3536m07{G\,353.6$-$0.7}
\def\gd087m01{G\,08.7$-$0.1}
\def\ge233m03{G\,23.3$-$0.3}
\def\pks{PKS$\;$2155$-$304}

\newcommand{\aj}{AJ}
\newcommand{\apj}{ApJ}
\newcommand{\apjl}{ApJL}
\newcommand{\apjs}{ApJS}
\newcommand{\aap}{A\&A}
\newcommand{\aapr}{A\&A Rev.}
\newcommand\araa{ARA\&A}
\newcommand\aaps{A\&AS}
\newcommand{\nat}{Nature}
\newcommand{\mnras}{MNRAS}
\newcommand{\prd}{Phys. Rev. D}
\newcommand{\pasj}{PASJ}

\newcommand{\HESSa}{HESS~J1427$-$608}
\newcommand{\HESSb}{HESS~J1626$-$490}
\newcommand{\HESSc}{HESS~J1702$-$420}
\newcommand{\HESSd}{HESS~J1708$-$410}
\newcommand{\HESSe}{HESS~J1731$-$347}
\newcommand{\HESSf}{HESS~J1841$-$055}
\newcommand{\HESSg}{HESS~J1857$+$026}
\newcommand{\HESSh}{HESS~J1858$+$020}

\newcommand{\HMS}[3]{$#1^{\mathrm{h}}#2^{\mathrm{m}}#3^{\mathrm{s}}$}
\newcommand{\DMS}[3]{$#1^\circ #2' #3''$}

\begin{frontmatter}


\title{Status of Very High Energy $\gamma$-ray Astronomy as of early 2008}
 \author{Arache Djannati-Ata\"i}
 \ead{djannati@apc.univ-paris7.fr}
\address{Astroparticule et Cosmologie (APC), CNRS, Universite Paris 7
  Denis Diderot, Paris, France\thanksref{label1}}
 \thanks[label1]{UMR 7164 (CNRS, Universit\'e Paris VII, CEA, Observatoire de Paris)}


\begin{abstract}
Data obtained in the very high energy $\gamma$-ray band with the new generation of imaging telescopes, in 
particular through the galactic plane survey undertaken by H.E.S.S., low
threshold observations with MAGIC and more recently by operation 
of VERITAS, have revealed dozens of galactic and extragalactic sources, providing a wealth 
of information on a variety of high energy acceleration sites in our universe. Also, the water Cherenkov
 instrument Milagro has provided several extended sources after seven years of data integration. An overview of
 these results with focus on some of the most recent highlights is given.
\end{abstract}

\end{frontmatter}
\section{Introduction}
\label{intro}
 Almost two decades after the establishment of the Crab nebula as the first TeV 
 emitting source, thanks to the pioneering work at the Whipple
 Observatory in Arizona~\cite{Weekes89}, the field of very high energy (VHE)
 $\gamma$-ray astronomy has entered the maturity age. 
 Although limited in aperture
 (3-6$\times10^{-3}$ sr, currently) and duty cycle ($\sim 10\%$), the
 imaging atmospheric Cherenkov telescopes (IACTs) have proved to be
 the most sensitive, thanks to precise angular reconstruction of
 the shower origin and to powerful background rejection
 capabilities. The new generation of these instruments 
 has allowed the discovery of more than 70 galactic and extragalactic
 sources. The major contributions have come from H.E.S.S. in
 Namibia (system of four 13~m diameter telescopes, 2003), MAGIC on La
 Palma (single 17 m diameter dish, 2004), CANGAROO-III in
 Woomera (three 10 m diameter dishes, one more pending, 2004), and  VERITAS (four 12 m diameter
 telescopes, 2006) in Arizona.
 Large aperture and duty cycle instruments, the
 water Cherenkov detector Milagro (1999-2007) and the particle counter array
 Tibet As-$\gamma$ (1992), although having a much lower
 sensitivity, have also interestingly
 contributed to the field through large exposures obtained after
 few-years scale integration times.
A brief overview of the field, as of early 2008, is given below.
 
\section{Galactic sources}
By late 2002, the TeV sky 
 consisted of only 6 confirmed sources, and was dominated by
 point-like extragalactic ones. Only one source, the Crab nebula,
 was firmly detected in the Galaxy.
A major breakthrough was accomplished by the
2004-2007 HESS Galactic Plane Survey (GPS): 
covering the inner Galaxy ($l \in[-85^{\circ}, 60^{\circ}], b \in[-2.5^{\circ}, 2.5^{\circ}]$), this survey has, up to
now, resulted in
the discovery of 40 galactic sources and the diffuse emission in the central 100 pc of the Milky Way,
while confirming 4 and invalidating 2 sources previously published by CANGAROO.
The Milagro sky survey has resulted in less prolific but nonetheless interesting discoveries: 
3 extended sources, 4 less significant hot-spots, and evidence for diffuse
emission along the Galactic Plane~\cite{Milagro:survey,Milagro:diffuse}.
When adding  the Crab, 3 discoveries\footnote{the detection of one source, Cyg$-$X1, still needs  
  confirmation, see table~\ref{tab:idsrces}.} and 2 confirmations made in the northern
hemisphere by MAGIC and VERITAS, the total number of known galactic
sources is well in excess of 50 (see Tables ~\ref{tab:idsrces},~\ref{tab:uidB} and \ref{tab:uidC}). 
In the following, different galactic  VHE source classes will be discussed with
focus on recent results.
\subsection{Supernova Remnants}
Following the original proposition of W. Baade and F. Zwicky back in
1934~\cite{BaadeZwicky34II}, 
but based on quantitative arguments (energetics, chemical composition) 
and diffusive shock acceleration
models (e.g. \cite{Ginzburg:1},~\cite{Blandford:1},~\cite{Drury83}), 
shell-type SNRs are considered as prime
acceleration sites for the galactic cosmic-rays, at most up to the
\emph{knee} ($\sim 10^{15}$~eV) (\cite{LagageCesarsky83}).   
The first signature of acceleration of cosmic electrons within shell-type
SNR shocks dates back to 1995~\cite{Asca:SN1006}
when ASCA observations of \sn1006 rims established the non-thermal
nature of their X-ray emission.     
$\gamma$-ray emission from SNRs, as a signature of interactions of
accelerated ionic cosmic-rays with internal or nearby matter have long been sought after by
space- and ground-based instruments.
VHE $\gamma$-rays were first detected from {\rxj1713}
by the CANGAROO~\cite{Cangaroo:rxj1713}
collaboration. The confirmation of  
this signal by HESS was made through the realization of the first
ever resolved image in the $\gamma$-ray band~\cite{Hess:rxj1713p1}. 
The latter image exhibited a clear shell morphology, strongly correlated with
non-thermal X-rays. 
 \begin{figure}
 \begin{center}
\includegraphics*[width=0.45\textwidth]{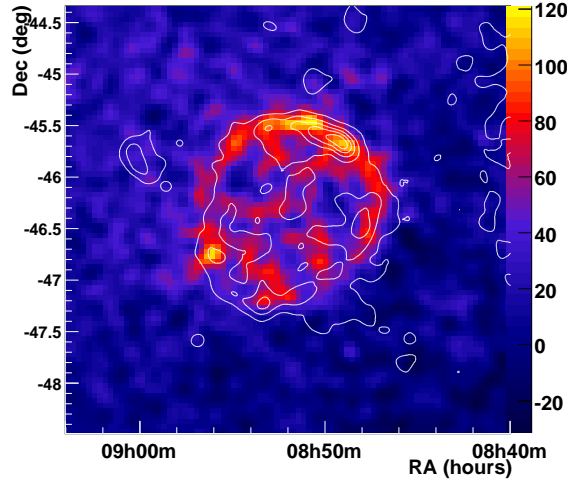}
  \caption{Smoothed VHE image of {\velajr} obtained by HESS, 
shown together with X-ray contours from the ROSAT All Sky Survey $>$\,1.3 keV~\cite{Hess:velajnrp2}.}
 \label{fig:velajr}
 \end{center}
 \end{figure}
$\gamma$-ray emission processes at play in {\rxj1713}, whether
leptonic or hadronic, have been and are still under
debate. Some of the pros and cons in each case are discussed
briefly hereafter. 
The $\gamma$- to X-ray correlation
tends to favour leptonic models, but that is at the
expense of a rather low
magnetic field strength B\,$\sim 10\,\mu$G, lower than
what is required by models of dynamical field amplification (see e.g.~\cite{BellLucek01}) in
non-linear shocks in order to explain the observed X-ray filamentary
features, i.e.  B\,$\sim 58-100\,\mu$G ~\cite{Voelketal2005}.
Hadronic scenarios also face difficulties --e.g. the required mean
pre-shock hydrogen number
density n\,$\sim\,$1\,cm$^{-3}$ violates the upper limit
n\,$<$\,0.02\,cm$^{-3}$ implied by the absence of thermal
X-rays~\cite{Cassam2004}-- and have recourse to quite low e$^-$/p
ratios~\cite {KatzWaxman08}, but they seem to better fit 
the shape of the VHE spectrum~\cite{Hess:rxj1713p2}. On the other hand
a detailed modeling of the interstellar
radiation field for the calculation of the inverse Compton (IC)
spectrum~\cite{Porter06} improves the fit to leptonic
models, though not for the latest published
spectrum~\cite{Hess:rxj1713p3}.
The uncertainties on age (1-10 kyr) and distance (1-6 kpc) leave also some
margin for different models; hence, for the time being, no
clear-cut distinction between them can be made.
The situation is analog for
the two other spatially resolved $\gamma$-ray SNRs, {\velajr} and {\rcwa86},
recently reported by
HESS~\cite{Hess:rcw86,Hess:velajnrp2}. The former shows also a strong correlation with non-thermal
X-rays and an absence of thermal X-rays, but has a
thinner VHE shell morphology (Fig.~\ref{fig:velajr}). {\rcwa86} is at variance with the two other TeV shell-type SNRs
in that it exhibits both thermal and non-thermal X-rays.

In the case of the very young and radio-bright SNR,
{\casa} --point-like in  $\gamma$-rays, detected initially
by the HEGRA collaboration~\cite{Hegra:casa} and confirmed
recently by MAGIC~\cite{Magic:casa}-- it 
seems again difficult to determine unambiguously the nature/loci of the emitting
particles with the current $\gamma$-ray data. Neutrinos, expected as secondary products of cosmic-ray
interactions with ambient matter, could be used to probe decisively  the hadronic
component of cosmic accelerators (see e.g. ~\cite{neutrinos:Becker08}). 
Estimation of event rates for the brightest $\gamma$-ray
SNRs, based on their VHE flux measurements, seems, however, to
imply the necessity of very
large volume detectors: indeed the 
potential signal to noise ratios for a 1\,km$^3$
class neutrino detector seem to be low, e.g. the expected signal and background rates for
{\rxj1713}, for an integration time of 5 years, are of order of 14 and 41,
respectively~\cite{neutrinos:kappes07}. 

VHE $\gamma$-ray observations of SNRs interacting with high density (i.e. n\,$>\,$10$^{3}$
\,cm$^{-3}$) molecular clouds in their vicinity are an alternative
probe of ionic acceleration by SNR shock waves. In this regard, older SNRs (i.e. with
age\,$>\,$few 10$^{4}$\,yr) are potentially attractive targets
since accelerated electrons must have lost much of their energy through
radiative cooling and should not reach multi-TeV energies~\cite{Yamazaki06}.
The VHE loci of W41/HESS\,J1834-026~\cite{Hess:scanpaper2,Magic:w41},
{\ic443}/MAGIC\,J0616+225~\cite{Magic:ic443}, also reported by VERITAS~\cite{Veritas:ic443_icrc},  and
W\,28/\rm\{HESS\,J1800-240, HESS\,J1801-233\rm\}~\cite{Hess:w28_icrc}, 
are coincident with such molecular clouds and suggest that
these objects are accelerating ionic cosmic-rays. The presence of OH
masers (tracers of shocked molecular matter) in {\ic443} and the
northeastern region of W\,28 supports this hypothesis.  
The VHE emission reported recently from another SNR/molecular cloud association
with a much younger object, CTB37A/HESS\,J1714-385~\cite{Hess:ctb37a}
enhances further this class of possible
hadronic accelerators, although, a PWN-type contribution
is also possible due to the discovery of an extended non-thermal
X-ray source near the VHE peak.
Another new candidate for this class is the formerly 'dark' source (see below),
HESS\,J1731$-$347, recently identified with a $\sim$\,30 kyr old SNR,
G353.6$-$0.7~\cite{Tian:j1731}. 

\begin{table}[tp]
\begin{center}
\caption{\label{tab:idsrces}Class A: Galactic sources with a firmly established counterpart
and for which the VHE emission origin/morphology (not necessarily
the emission process) is also fairly well identified; see text.}
\smallskip
{\linespread{.8}
\footnotesize\rm
\begin{tabular*}{0.48\textwidth}{@{\extracolsep{\fill}}lllllllll}
\hline\hline
VHE Class    & Object           & discovery$^a$ \\
\hline
Shell        & \velajr          & CANGAROO~\cite{Cangaroo:velajnr,Hess:velajnrp1,Hess:velajnrp2}\\  
Shell        & \rxj1713         & CANGAROO \cite{Cangaroo:rxj1713,Hess:rxj1713p1,Hess:rxj1713p2}\\
Shell        &\rcwa86           & HESS~\cite{Hess:rcw86}    \\
SNR          &\casa             & HEGRA \cite{Hegra:casa}   \\ 
PWN          &Crab nebula       & Whipple\cite{Weekes89,Themistocle:crab,Whipple:crabspec}\\
PWN          &\ga09p01          & HESS~\cite{Hess:g09}\\
PWN          &MSH\,15$-$52      & HESS (J1514$-$591)~\cite{Hess:msh1552}\\
PWN          &Vela\, X          & HESS (J0835$-$455)~\cite{Hess:velaxlett}\\
PWN          &G\,18.0$-$0.7     & HESS (J1825$-$137)~\cite{Hess:scanpaper1,Hess:scanpaper2}\\
PWN          &K3/Kookaburra     & HESS (J1420$-$607)~\cite{Hess:kookaburra} \\
PWN          &G\,21.5-0.9       & HESS~\cite{Hess:youngpwneicrc}\\
PWN          &PSR~J1718$-$3825  & HESS (J1718$-$385)~\cite{Hess:j1718j1809}           \\
PWN          &PSR~J1809$-$1917  & HESS (J1809$-$193)~\cite{Hess:j1718j1809}           \\
PWN$^\dag$   &Kes\,75           & HESS~\cite{Hess:youngpwneicrc}\\
Binary       &PSR~B1259$-$63    & HESS~\cite{Hess:psrb1259} \\
Binary       &LS~5039           & HESS~\cite{Hess:ls5039p1,Hess:ls5039p2}\\
Binary       &LSI~+61~303       & MAGIC~\cite{Magic:lsi61,Veritas:lsi61}\\
Binary       &Cyg$-$X1$^\ddag$    & MAGIC~\cite{Magic:cygx1}\\
\hline\hline
\end{tabular*}
\begin{flushleft}
\smallskip
$^a$For extended PWNe the best fitted position of the source is quoted as
well. Additional references to the discovery paper are given when
relevant, e.g., confirmation of the source or discovery of
important features (morphology, spectrum).\\
$^\dag$Contribution from the SNR shell is not excluded for Kes~75.\\
$^\ddag$The firm detection of this source is not established yet.\\
\end{flushleft}
}
\end{center}
\end{table}

\begin{table}[tp]
\begin{center}
\caption{\label{tab:uidB}Class B: Galactic sources with either an
  identified or a plausible counterpart but 
for which further data is required to firmly establish the association
or type of emission; see text. The VHE types in last column are tentative scenarios put forward
by different authors and are not
exclusive of other possibilities; MoC-SNR stands for molecular
cloud-SNR associations; OpC: open cluster; Lept.: leptonic}
\smallskip
{
\linespread{.8}
\footnotesize\rm
\begin{tabular*}{0.48\textwidth}{@{\extracolsep{\fill}}lll}
\hline\hline
Object$^a$                   &Poss. counterpart(s)   &Type                   \\
\hline
MAGIC\,J0616+225$^\S$        &  \ic443             &    MoC-SNR~\cite{Magic:ic443}               \\
HESS\,J1023$-$575            &  Westerlund2        &    OpC wind        \\
HESS\,J1303$-$631            &  PSR~J1303$-$6315   &    PWN        \\
HESS\,J1357$-$645            & PSR\,J1357$-$6429          &    PWN              \\
HESS\,J1418$-$609            &  G\,313.3+0.1       &    PWN                \\
HESS\,J1616$-$508$^\dag$      & PSR\,J1617$-$5055   &    PWN/SNR           \\
HESS\,J1640$-$465 $^\S$      &  G\,338.3$-$0.0     &    PWN/SNR           \\
HESS\,J1702$-$420            & PSR\,J1702$-$4128          &    PWN              \\
HESS\,J1713$-$381$^\S$       &  \ctbb37            &    SNR      \\
HESS\,J1714$-$385$^\S$        &  \ctba37            &    PWN/MoC-SNR              \\
HESS\,J1731$-$347$^\dag$$^\S$ &  \gc3536m07         &    SNR-Shell          \\
HESS\,J1745$-$290              &  SgrA$^{\star}$/G\,359.95$-$0.04      &    BH/PWN        \\
HESS\,J1804$-$216$^\S$        &  G\,08.7-0.1        &    PWN/SNR         \\
HESS\,J1813$-$178$^\dag$$^\S$  &  \gb128m02          &    PWN/SNR       \\
HESS\,J1800$-$240$^\ddag$$^\S$ &  \w28 south          &    MoC-SNR        \\
HESS\,J1801$-$233$^\S$        &  \w28 north-east       &    MoC-SNR       \\
HESS\,J1834$-$026$^\S$        &  W41                &    MoC-SNR        \\
HESS\,J1837$-$069$^\dag$      &  PSR\,J1838$-$0655     &    PWN              \\ 
HESS\,J1857+026$^\dag$        & PSR\,J1856+0245        &   PWN              \\
HESS\,J1912+101              & PSR\,J1913+1011        &    PWN              \\
TeV~~J2032+4130$^\dag$        & extended X-ray  &    Pevatron/Lept.~\cite{horns:tev2032}\\
\hline\hline
\end{tabular*}
\begin{flushleft}
\smallskip
$^a$References for the HESS sources are available at
{\tt http://}\\{\tt www.mpi-hd.mpg.de/hfm/HESS/public/HESS\_catalog.htm}; \\
$^\dag$These sources were initially classified as dark.\\
$^\S$These sources have a rather well identified counterpart, except that the emission
origin/type is still ambiguous.\\
$^\ddag$This source splits in three sub-sources, J1800$-$240 A, B
and C; the former two coincide with molecular matter. \\
\end{flushleft}
}
\end{center}
\end{table}
\subsection{Pulsar Wind Nebulae (PWNe)}

Relativistic particles of the shocked winds of pulsars
shine through synchrotron and IC radiation from radio to
$\gamma$-rays and form Plerions (\emph{Pleres
  Plera} or filled bags).  
The \emph{Chandra} and {\it XMM-Newton} imaging and spatially resolved spectroscopy of X-ray synchrotron nebulae 
have provided a wealth of details on pulsar winds and their
interactions with the medium (e.g. ~\cite{PWNReviewKargaltsev08}). Two morphological
types have emerged:  
those which show a toroidal structure around the pulsar, with one or two jets 
along the torus axis, and those dominated by a cometary structure, 
with the pulsar close to the comet apex. Also, the spectral softening of the extended
emission as a function of distance from the pulsar, observed for many PWNe,  
has been successfully interpreted as the synchrotron cooling of the
X-ray emitting electrons. 

The first source discovered in the VHE domain, the Crab nebula, is a
plerion and still exhibits a point-like emission to the precision of the current
instruments (few arc-minutes).
As was remarkably soon predicted by~\cite{Gould65},
the study of its synchrotron and IC components opened the
way to the measurement of the magnetic field strength within the
nebula~\cite{deJagerHarding92}. The HESS GPS has revealed a large
number of PWNe, most of which are very extended and associated with
energetic and middle-aged pulsars, with age $\sim10^4-10^5$ yr.  
HESS\,J1825$-$137 can be considered as the prototype of such objects:
its $\gamma$-ray
emission is extended, with a characteristic size of 0.3$^{\circ}$, and
offset to the south of the pulsar 
B\,1823$-$13; the latter has spin-down characteristic age
 and power of $\tau_C=$\,21000 yr and $\dot{E}=10^{35}$ erg\,s$^{-1}$,
 respectively. It is remarkable that the X-ray nebula exhibits exactly 
the same feature and is offset to the south of the pulsar, except
that its extension is of the order of a few arc-minutes rather than a
fraction of a degree. This can be naturally explained in terms of the
cooling time of particles in the estimated average nebular magnetic field B\,$\sim 10\,\mu$G: 
the X-ray generating electrons have higher energies than those responsible via IC 
for the VHE $\gamma$-rays, hence they cool faster and have a
shorter range, whereas the latter include relic particles, i.e. those
injected and accelerated at early epochs of the pulsar
activity~\cite{Hess:j1825p1}. This interpretation has been further
supported by the discovery of the spectral softening of
the VHE nebula as a function of the distance from the pulsar, analogous to that seen in X-rays~\cite{Hess:j1825p2}.
The offset of the nebula with respect to the pulsar can be understood
in terms of the displacement caused by an anisotropic reverse shock, itself 
due to the explosion of the progenitor in an inhomogeneous medium.
This explanation was
first proposed by ~\cite{Chevalier98} for Vela\,X, another
asymmetrical PWN in radio and X-rays, which is, as has been
demonstrated by HESS~\cite{Hess:velaxlett}, also offset in VHE $\gamma$-rays.
A number of other extended offset nebulae have been
discovered by HESS and the systematic search for molecular clouds in
their vicinity \cite{Lemiere06} has revealed in many cases clouds at compatible
kinematic distances to their associated pulsar, as candidates to
explain their peculiar morphology.
Another key point in the study of these middle-aged nebulae is the possibility of access
both to their 'current' state through X-rays (fresh short-lived
particles), and to their past history and evolution, e.g. those of the
pulsar and magnetic field, through the the VHE emission (relic
electrons)~\cite{Lemiere06,deJagerDjannati08}.
  
In this context the young Crab nebula appeared as a rather particular
VHE source\footnote{this was already the case at other wavelengths}. 
Very recently, two other very young nebulae were discovered by HESS: 
G\,21.5$-$0.9 which harbors the second most energetic pulsar known in
the Galaxy (after the Crab) and Kes\,75, associated with the 325\,ms,
X-ray only,  pulsar, PSR\,J1846-258~\cite{Hess:youngpwneicrc}. 
Despite their similar young ages to the Crab nebula, G\,21.5$-$0.9 and
Kes\,75 exhibit much
smaller X-ray to $\gamma$-ray luminosity ratios and hence a much lower
nebular field. Also both objects are 
classified as composite SNRs and, as such, the possibility of VHE
radiation from particles accelerated at the forward shock of the
freely expanding shells should be considered. While this possibility remains open
for Kes\,75, the low gas densities inferred through thermal X-ray
measurements for G\,21.5$-$0.9 make the contribution of the SNR shell to its
$\gamma$-ray emission unlikely.  

\begin{table}[bh]
  \begin{center}
{
\linespread{.8}
\footnotesize\rm
    \begin{tabular*}{0.48\textwidth}{@{\extracolsep{\fill}}ll}
      \hline\hline
      Source & Comments                \\
      \hline
      HESS~J0632+057 & point-like, near Monoceros Loop~\cite{Hess:monocerosicrc}\\
      \HESSa                  &  ext., PWN compatible~\cite{Hess:uid}\\
      HESS~J1614$-$518 & ext., soft~\cite{Hess:uid}\\
      \HESSb                  &  ext. hard ~\cite{Hess:scanpaper2}\\
      HESS~J1632$-$478 &     ext., IGR srce, no MOR ~\cite{Hess:scanpaper2}  \\ 
      HESS~J1634$-$472        & ext., soft, SNR, no MOR ~\cite{Hess:scanpaper2}\\ 
      \HESSd           &      ext., PWN compatible ~\cite{Hess:scanpaper2}\\
      HESS~J1745$-$303        & ext., MoC-SNR to the north~\cite{Hess:scanpaper2,Hess:j1745int}\\ 
      \HESSf                  & ext.~\cite{Hess:uid} \\
      \HESSh                  & ext.~\cite{Hess:uid}  \\
      MGRO~J1908+06           & HESS J1908+063/GRO srce~\cite{Hess:j1908icrc,Milagro:survey}\\
      MGRO~J2019+37           & ext. $>$\,1$^\circ$ ~\cite{Milagro:survey}\\
      MGRO~J2031+41           & ext.~\cite{Milagro:survey}\\
      \hline\hline     
    \end{tabular*}
}
    \caption{\label{tab:uidC}Class C: Galactic sources for which
      no clear counterpart exists at other wavelengths: 'ext.' stands
      for extended, hard means a photon spectral index
      $\Gamma\,<\,$2.2 and 'no MOR' is to indicate the morphological
      incompatibility of the source with lower-wavelength objects, if any, in its line
      of site.
}
  \end{center}
\end{table}

\subsection{$\gamma$-ray binaries}
Although a plethora of binary systems are X-ray emitters,
only three objects have been firmly detected in the VHE band up to now:
PSR\,B1259$-$63/SS\,2883, LS\,5039 both reported by HESS and
LSI\,+61\,303, discovered by MAGIC~\cite{Magic:lsi61}.
PSR\,B1259$-$63 is a 48 ms radio pulsar,  but for LS\,5039 and  LSI\,+61\,303 
the precise nature of the compact object is not known: the 4
$M_{\odot}$ upper limit on their mass is consistent both with neutron
stars and low mass black holes. These two systems are much closer
binary systems with periods of 3.9 and 26.5 days, respectively, as
compared to the 3.4 yr period of PSR\,B1259$-$63/SS\,2883.

The VHE emission of LS\,5039 is clearly periodic, with enhanced
and harder emission at the inferior conjunction.  The variability of
LSI\,+61\,303 has been recently confirmed by
VERITAS~\cite{Veritas:lsi61}, but it is not
clear yet if its emission is strictly periodic.
The main question regarding these sources is whether the relativistic particles come from
accretion-powered jets or from a rotation-powered pulsar wind as 
for PSR\,B1259$-$63/SS\,2883.
Also, although the interaction of the relativistic wind of the latter 
should play a major role, the precise emission mechanism, whether leptonic or hadronic, is unknown
(see e.g.~\cite{Dubus06}).

The marginal detection of Cyg$-$X1, a $>$13$M_{\odot}$ black hole
binary system, recently reported by
MAGIC~\cite{Magic:cygx1}, if confirmed, is interesting in this context,
since it should require rather an accretion-powered jet model. 
\subsection{Unidentified Sources}
It is remarkable that most of the galactic sources are extended and many of
them show a featureless morphology ; this precisely renders their
identification difficult, except when a clear correlation
with an object at other wavelengths exists, and/or a coherent
multi-wavelength model is found. Finding the counterpart
of a point-like source is, in principle,
straightforward (when it exists), but the identification of the VHE
emission process requires again a coherent model. Given these
boundary conditions, the following classes can be defined:\\ 
A) Sources with a firmly established counterpart
and for which the VHE emission origin/morphology (not necessarily
the emission process) is also fairly well identified, e.g the Crab
nebula, TeV shell-type SNRs and
$\gamma$-ray binaries fall into this class.\\
B) Sources with either an identified or a plausible counterpart but 
for which further data is required to firmly establish the association
or the type of emission, e.g., PWN- and/or SNR-type.\\ 
C) Sources with no plausible counterparts (or 'dark' sources).\\

The majority of the galactic sources fall into the last two classes
(Tables~\ref{tab:uidB} and \ref{tab:uidC}) and hence many are still
considered as unidentified. 
TeV\,2032+4130, the first unidentified source discovered by HEGRA~\cite{Hegra:tev2032} in 2002,
has been confirmed by
MAGIC~\cite{Magic:tev2032}. Recently an extended X-ray source matching the
position of TeV~J2032+413 was detected through deep {\it XMM-Newton}
observations~\cite{horns:tev2032} and consequently this source is no more
considered as a dark one, but is classified as B. This is the case
also for five HESS sources, which were previously classified as dark:
HESS\,J1731$-$347 has now an old shell-type SNR counterpart whereas a
young SNR has been discovered and associated with HESS\,J1813$-$178; in addition two young
and energetic pulsars have been discovered in the vicinity (line of
sight) of HESS\,J1837$-$069 and HESS\,J1857+026, and HESS\,J1616$-$508
has now a faint X-ray counterpart.
Among the three unidentified sources reported by
Milagro~\cite{Milagro:survey}, MGRO\,J2019+37 and MGRO\,J1908+06 have been recently confirmed 
by Tibet As-$\gamma$~\cite{Tibet:mgro2019} and
HESS~\cite{Hess:j1908icrc}, respectively, but remain still unidentified. 

The dark sources are prime hadronic accelerator candidates. However,
as noted first by ~\cite{Lemiere06,deJagerDjannati08}, due to the
large lifetime of VHE 
emitting electrons (up to a few 10\,$\times$\,kyr depending on the
nebular field) the ratio of the X-ray luminosity to the $\gamma$-ray
luminosity is a decreasing function of the system age and hence one expects TeV PWNe with such
faint X-ray counterparts that they could well be below the sensitivity
threshold of current X-ray instruments. Hence it is likely that some of the dark
sources are indeed ``$\gamma$-ray PWNe'' without multi-wavelength counterparts. 

It is also noteworthy that the extended class B source HESS\,J1023$-$575 is in
coincidence with the second most massive young cluster in the Milky
Way, Westerlund\,2. Strong shocks created through the colliding winds
of massive stars are believed to be able to accelerate particles up to
TeV energies and their collective effects can in principle provide
sufficient energy for the observed emission. If so, a new class of
cosmic ray accelerators should emerge through observations of other
clusters of this type.

\subsection{Galactic Center and its neighbourhood}
The point-like emission from the direction of the Sgr~A complex was discovered
already 4 years ago by CANGAROO \cite{GC:CANGAROO}, and subsequently
detected by Whipple \cite{GC:Whipple04},  and by
HESS \cite{GC:HESS}; the latter has produced the
most consistent measurements of its signal and spectrum. Due to the variety of
potential TeV emitting sources --including the massive black hole
Sgr~A$^{\star}$-- the identification of the Galactic Center TeV emission 
origin is a difficult task. Two paths have been followed by HESS:
simultaneous observations of X-ray flares of Sgr~A$^{\star}$ with
\emph{Chandra} and the improvement of the source (HESS\,J1745$-$290) localisation.
The former approach has resulted in an upper-limit on the flaring TeV
component with respect to the steady emission during the
observed X-ray flare~\cite{463} , while the latter has allowed HESS to
constrain the source localisation with comparable and 
extremely low statistical and systematic errors --better than $6''$.
This precision is enough to exclude the SNR Sgr\,A\,East 
as the dominant source of the TeV emission, leaving the
PWN candidate G\,359.95$-$0.04 and  Sgr~A$^{\star}$
as the most likely counterparts~\cite{286}. The dark matter
interpretation is also clearly disfavoured: the measured power-law
spectrum seems quite incompatible with typical (quark or gluon-fragmentation type) neutralino
annihilation scenarios~\cite{Hess:darkmatter06}.   

HESS\,J1745$-$290 lies above a diffuse emission along the
Galactic ridge. HESS data have shown a clear correlation with the giant molecular clouds 
of the central $\sim$100~pc of the Galaxy~\cite{Hess:gc_diffuse}, and a
spectrum which is harder (index of 2.3) than that expected for
$\gamma$-rays, if they were produced through interactions of cosmic rays
with the same spectrum as the one local to the solar
system. One elegant explanation for this is the reduced effects of diffusion and
escape due to the proximity of accelerators and
targets.
Another feature of the $\gamma$-ray emission is its deficit 
as compared to the density of molecular clouds
around $l\simeq$1.5$^\circ$. This has been interpreted in terms of a time
limited diffusion range of the cosmic rays under the assumption that
they were accelerated only 'recently' (some 10000
years ago) near the Galactic Center~\cite{Hess:gc_diffuse}, but the
question remains open to alternative explanations such as assuming 
that the $\gamma$-ray emission is a superposition of point-like
sources distributed according to the distribution of the molecular gas.

\section{Extragalactic Sources}

\begin{table}
\caption{\label{tab:agn}VHE $\gamma$-ray emitting AGN, as of early
2008, ordered by redshift.
The last column gives the reference to the discovery
publication and, when relevant, to the confirmation papers. 
}
\smallskip
{\linespread{.8}
\footnotesize\rm
\begin{tabular*}{0.48\textwidth}{@{\extracolsep{\fill}}llllll} \hline\hline 
Object        &         $z$     &  Class &   Discovery        &  Ref. \\\hline
M\,87         &         0.004   &       FRI     &  HEGRA    2003  & \cite{Hegra:m87,Hess:m87} \\
Mrk\,421      &         0.031   &       HBL     &  Whipple  1992  & \cite{Whipple:mrk421, Whipple:mrk421spec}     \\
Mrk\,501      &         0.034   &       HBL     &  Whipple  1996  & \cite{Whipple:mrk501, Hegra:mrk501spec}  \\
1ES\,2344+514 &         0.044   &       HBL     &  Whipple  1998  & \cite{Whipple:1es2344, Whipple:1es2344spec}	      \\
Mrk\,180      &         0.046   &       HBL     &  MAGIC    2006  & \cite{Magic:mrk180}	      \\
1ES\,1959+650 &         0.047   &       HBL     &  TA       2002  & \cite{TA:1es1959, Whipple:1es1959, Hegra:1es1959}   \\
BL\,Lac       &         0.069   &       LBL     &  MAGIC    2006  & \cite{Magic:bllac}   \\
PKS\,0548-322 &         0.069   &       HBL     &  HESS     2006  & \cite{Hess:pks0548_icrc} 	      \\
PKS\,2005-489 &         0.071   &       HBL     &  HESS     2005  &\cite{Hess:pks2005} 	      \\
RGB\,0152+017 &         0.080   &       HBL     &  HESS     2008  &\cite{Hess:rgb0152} 	      \\ 
W\,Comae      &         0.102   &       IBL     &  VERITAS  2008  &\cite{Whipple:wcomae}	      \\
PKS 2155-304  &         0.116   &       HBL     &  Durham   1999  &\cite{Durham:pks2155,Hess:pks2155}\\
H\,1426+428   &         0.129   &       HBL     &  Whipple  2002  &\cite{Whipple:h1426,Hegra:h1426,Cat:h1426}  \\
1ES\,0806+524 &         0.138   &       HBL     &  VERITAS  2008  &\cite{Whipple:1es0806}	      \\
1ES\,0229+200 &         0.140   &       HBL     &  HESS     2007  &\cite{Hess:1es0229} 	      \\
H\,2356-309   &         0.165   &       HBL     &  HESS     2005  &\cite{Hess:ebl}	      \\
1ES\,1218+304 &         0.182   &       HBL     &  MAGIC    2005  &\cite{Magic:1es1218}	      \\
1ES\,1101-232 &         0.186   &       HBL     &  HESS     2005  &\cite{Hess:ebl}	      \\
1ES\,0347-121 &         0.188   &       HBL     &  HESS     2007  &\cite{Hess:1es0347} 	      \\
1ES\,1011+496 &         0.212   &       HBL     &  MAGIC    2007  &\cite{Magic:1es1011} \\
PG\,1553+113  &       $>0.25$   &       HBL     &  HESS     2005  & \cite{Hess:pg1553} \\
3C\,279       &         0.536   &      FSRQ     &  MAGIC    2007  & \cite{Magic:3c279icrc}     \\
S5\,0716+71   &         unknown &       HBL     &  MAGIC    2008  &\cite{Magic:s50716} \\
\hline\hline                         
\end{tabular*}
}
\end{table}

The first VHE emitting extragalactic source, Mrk~421, was discovered back
in 1992~\cite{Whipple:mrk421}.
There has been tremendous progress in this area since 2003 and, as of
early 2008, 23 extragalactic sources are known to be VHE
$\gamma$-ray emitters. All of these sources but one are blazars, 
i.e. belong to the class of radio-loud Active Galactic Nuclei (AGN) with one of
radio jets pointing towards the observer at small angles ($\sim$ few degrees).   
The broad band spectra of blazars are characterized by two 
broad peaks, in mm$-$soft X-rays and MeV$-$GeV bands, respectively. 
The lower energy peak is understood as synchrotron emission of energetic leptons within
the relativistic jet, and the generally agreed upon origin for the second
component is IC scattering of either synchrotron photons
(SSC)\footnote{for Synchrotron Self-Compton} or
ambient photons (EC)\footnote{for External Compton} by the same
population of leptons. Alternatively,
hadronic models are put forward for the higher energy component;
however the observed strong correlations between the X-ray and the
$\gamma$-ray emissions favour rather leptonic models\footnote{There exists
however a noteworthy exception in the history of the field: a TeV flare without any counterpart in
X-rays was detected during observations of 1ES\,1959+650 on June 4,
2002~\cite{Whipple:1959orphan}.}.

As listed in Table~\ref{tab:agn}, the High frequency peaked BL Lac objects, or
HBLs, i.e. those for which the lower energy component peaks in X-rays,
are the most prominent  TeV emitting blazars. The three exceptions are: BL Lac
itself, classified as an LBL (Low freq. BL Lac), W\,Comae, an
Intermediate BL Lac object, and 3C~279 a flat spectrum radio quasar,
or FSRQ.

M~87, the well-known nearby FRI radio-galaxy, is the
first non-blazar source and its detection is of
particular interest. 
Its two-day variability time scale, first measured by
HESS~\cite{Hess:m87} and
recently confirmed by VERITAS~\cite{Veritas:m87icrc}, constrains
the size of the emission region $\sim$\,5$\delta$\,R$_{\rm
  s}$, dramatically close to that of the
black hole Schwarzschild radius R$_{\rm s}$, the expected Doppler
factor $\delta$ being quite small given  
the large declination angle of the M~87 jet to the line of sight ($\sim$~30$^\circ$).

There have been two recent highlights in blazar
observations: 1) the very fast variability of {\pks} observed by HESS
during a dramatic flaring episode in July 2006; the best measured 
individual flare rise time is of 173$\pm$23
seconds~\cite{Hess:pks21551stflare} and implies, within one-zone
SSC models when using causality, a huge Doppler
factor of order 100 which is in conflict with those
deduced from the unification models between blazars and
radio galaxies\cite{henrisauge:bulklorentzcrisis}; this requires the development of
inhomogeneous models; 2) evidence reported by MAGIC for time lag between high- and
low-energy band photons during 2 flares of Mrk~501, which may be an indication of
progressive acceleration of leptons within the jet~\cite{Magic:mkn501lag}. 

Beyond the understanding of blazars themselves, measurement of their
VHE spectrum and its attenuation through pair creation due to Extragalactic Background Light (EBL)
can be used to constrain the EBL density itself and, thereby, the star
formation history of the universe. 
The most recent highlight is the discovery of VHE $\gamma$-rays from
3C~279 by MAGIC at z = 0.536. The HESS detections of
hard spectra from 1ES~1101$-$232 (z = 0.186) and H~2356$-$309 (z =
0.165) implied already a low 
level of EBL in the optical/near-infrared wavelengths~\cite{Hess:ebl},
very close to the lower limit given by the integrated light of
resolved galaxies. The detection of 3C~279
represents a major step in redshift and should put severe limits in the
sub-micron to 2$\mu$ band. It is remarkable that the possibility of
constraining the EBL through $\gamma$-ray measurements was
predicted more than 15 years ago following the detection by Egret of
the same source, 3C~279~\cite{SteckerJagerSalomon:ebl3c279}.

\section{Summary}

The field of VHE $\gamma$-ray astrophysics has gone through a dramatic evolution
since 2004, thanks to the high sensitivity of the new generation
IACTs. The HESS GPS represents a major step in that it has revealed, beyond
the large number of sources, diverse classes of $\gamma$-ray emitting
galactic objects and acceleration sites: young shell-type SNRs,
SNRs interacting with molecular clouds, middle-aged offset PWNe, very young
composite PWNe and $\gamma$-ray binaries. Given the large number of still
unidentified sources, other potential classes of sources could emerge,
including the promising case of massive star clusters.
The increasing number of blazar sources in the extragalactic domain
allows now for population studies, and one non-blazar source, M~87 is
under scrutiny, in particular  by VERITAS. Also, while the early attempts to constrain the
intergalactic radiation field suffered from the very limited number of
sources and a reduced range in redshift, the growing number
of objects, and especially the detection of 3C~279 obtained at a low
energy threshold by MAGIC, have definitely opened the path towards the cosmological
application of $\gamma$-ray astrophysics.
There is no doubt that VHE $\gamma$-ray  astronomy is now a genuine branch of astronomy with multiple
connections to cosmology and fundamental physics.



\end{document}